\documentclass{jetpl}
\twocolumn

\lat

\DeclareMathOperator{\ol}{\widehat{L}}
\DeclareMathOperator{\os}{\widehat{S}}
\DeclareMathOperator{\ok}{\widehat{K}}
\DeclareMathOperator{\oc}{\widehat{C}}
\DeclareMathOperator{\oev}{\widehat{E}} 

\DeclareMathOperator{\wtn}{\widetilde{\nu} } 
\DeclareMathOperator{\oln}{\overline{\nu} }
\DeclareMathOperator{\grp}{\nabla_{\! P}} 

\title{On true relaxation statistics in gases}

\rtitle{On true relaxation statistics in gases}

\sodtitle{On true relaxation statistics in gases}

\author{Yu.\,E.\,Kuzovlev\/\thanks{kuzovlev@kinetic.ac.donetsk.ua, yuk-137@yandex.ru}}

\rauthor{Yu.\,E.\,Kuzovlev}

\sodauthor{Kuzovlev}

\address{Donetsk Free Statistical Physics Laboratory}

\abstract{By example of a particle interacting with ideal gas, 
it is shown that statistics of collisions in statistical mechanics 
at any degree of the gas rarefaction qualitatively differs from 
that conjugated with Boltzmann's hypothetical molecular chaos and kinetic equation. 
In reality, probability of the particle collisions in itself is random, 
which results in power-law asymptotic of the particle velocity relaxation. 
An estimate of its exponent is suggested basing on simple 
kinematic reasonings.}


\begin{document}

\maketitle

{\bf 1}.  In the classical Boltzmann's picture of molecular chaos, 
inter-particle collisions are treated like momentary random events 
with concentration (number per unit space volume per unit time)  assumed   
in the Boltzmann equation (BE)~\cite{rl,bog} to be proportional to product  
of one-particle distribution functions (DF) of participating particles. 
In reality, collision is process absorbing relative motion of colliding particles. 
Therefore spatial distribution of concentration of collisions as fictitiously whole 
momentary events must drift with center-of-mass velocity of 
their participants. By this reason, as it was noticed in~\cite{i1}, 
the distribution can not reduce to product of one-particle DFs  
of the participants. 
 In other words, the Boltzman's molecular chaos hypothesis   
is incompatible with exact equations of mechanics and the related  BBGKY   
equations~\cite{rl,bog}. They do not allow to prescribe to collisions 
a definite probability. 

Importantly, such reasonings are indifferent to value of (mean) number 
density of particles, \,$\nu$\,, and remain valid in the ``Boltzmann-Grad limit'' (BGL) when  
 radius of  interaction between particles  decreases, $a\rightarrow 0$,  while 
their density increases, $\nu\rightarrow \infty$,  
in such manner that their characteristic mean free path length keeps constant,  
$\lambda=(\pi a^2\nu)^{-1} =\,$const\,,  although characteristic gas parameter (GP)    
turns to zero, $a^3\nu \propto a/\lambda \rightarrow 0$. 
 
To the same conclusion,  -   that BE is a model contradicting rigorous statistical  
mechanics, -   we can arrive  also in various other  
ways, -   see~\cite{ ufn,tmf,pufn,lufn} and references therein. 
At the same time, in ``mathematical physics'' formal 
substantiation of BE is constructed for at least 
physically rather meaningless limit    
\,$a\rightarrow 0$,  $N\rightarrow \infty$, $N a^2 =\,$const\,, 
with \,$N$  being total number of gas particles  \cite{lan,sr}  
(also called ``Boltzmann-Grad limit''). 
Therefore, in view of practical importance of the question~\cite{lufn,kr}, 
additional indications of principal BE's defects would be useful. 
All the more in view of today's use of idealized collision language    
in description of many-particle processes, e.g. in gas of ``partons'',  
even in the high-energy physics (see \cite{ct,ww}). 
 
Notice that our previous  ``letter'' to the present journal \cite{0710}  
was rejected with advise to search another place for our material. 
It became \cite{tmf},  and here we  consider a new approach 
to the kinetics of molecular Brownian motion, now focusing  on statistics 
of collisions themselves rather than spatial displacements 
of test particle in homogeneous gas. At that, for more pungency and clarity, 
we exclude interactions between gas particles (``atoms'') and thus any 
hydrodynamic effects. 

{\bf 2}. Just chosen system is particular case of two-component gas, with  
one of components being so  ``rare impurity'' that  
in visible space region it is represented by a single particle.   
In the Boltzmann kinetics, DF of coordinates and velocity of this particle,      
 $F_0(t,R,V)$\,, undergoes linearized BE, 
sometimes called also~\cite{rl}  Boltzmann-Lorentz equation (BLE), 
\begin{align}
\partial_t{F}_0 = [ -V\nabla + \nu \widehat{K} ]\, F_0 \, , \label{ble}
\end{align}
with~$\widehat{K}$ designating Boltzmann-Lorentz kinetic operator  (BLO)  
acting by formula  
\begin{align}
\ok F(V) =  \int_v  \int \!  d^2b\, \,
|v-V| \, [\oc -1] \, F(V) \, G(v) \, ,   \label{ci}  
\end{align}
where~\,$G(v)$\, is DF of velocity  \,$v$\, of atoms,  
\,$b$\, is the impact parameter vector ($b \perp v-V$), 
while \,$\int_v =\int d^3v$\, and \,$\oc =\oc(b,\dots)$ 
are collision integral and operator. 

  In rigorous statistical mechanics we have to start from the BBGKY hierarchy  
  \cite{rl,bog}.  
 For our system it under standard normalization~\cite{bog,tmf}    
 reads~\cite{tmf,0806} 
 \begin{align}
\partial_t{F}_n = [ -V\nabla + \sum_{k=1}^n \ol _k ] \,F_n - 
\nu\grp \!\! \int_{n+1} \! f_{n+1} \,F_{n+1} \, , \label{bge} 
\end{align}
where\, \,$\ol_k = -(v_k-V)\,\nabla_{\rho_k}  + f_k\,(\nabla_{\! p_k} - \nabla_{\! P})$\,   
for \,$k >0$\,,  \,$P=MV$ is momentum of our test ``Brownian'' particle (BP), 
\,$M$\, is its mass, \,$p_k=m v_k$ and $m$\, momenta and mass of atoms, 
\,$\rho_k = r_k-R$\, with \,$r_k$\, being atoms' coordinates,  
\,$f_k \hm = f(\rho_k)$\,, \,$f(\rho)=-\nabla_{\rho} \Phi(\rho)$\,, 
 \,$\Phi(\rho)$\, is BP-atom interaction potential (of course, repulsive, 
short-range and let spherically symmetric), 
and \,$\int_k = \int d^3v_k \int d^3\rho_k $\,. 
 If at initial time moment, let $t=0$, correlations between BP and gas  
are completely absent, then \,$F_n |_{t=0} = F_0(0,R,V) \prod_{k=1}^n G(v_k) $\,.  
Solution to equations (\ref{bge})  is easy obtainable by their direct consecutive 
time integrations: 
\begin{eqnarray}
F_0 = [ \,\os_0(t) -\nu \int_{t>t_1>0} \os_0(t-t_1)   \, \label{is}
\times \\  \times \, 
\nabla_{\! P} \!  \int_1 f_1 \os_{1}(t_1) \, G_1  \,+  \nonumber 
\\ + \, \nu^2 
\int_{t>t_1>t_2>0}  \os_0(t-t_1) \, \nabla_{\! P} \!\! \int_1 f_1  
\os_{1}(t_1-t_2)  \,\times \nonumber \\ \times 
\nabla_{\! P} \!\! \int_2 f_2  \os_{2}(t_2)  \,G_1 G_2 \, + \dots\,] 
\, F_0|_{t=0} \,    \nonumber 
\end{eqnarray}
with designations \,$\os_{n}(\tau) =\exp{ \{\,\tau \,[\,-V\nabla +  
\sum_{k=1}^n \ol _k \,]\} }$\,  and \,$G_k= G(v_k)$\,.  
 But summation of this series is significant problem. 

 {\bf 3}.  It may be thought to become simpler when GP is small (in BGL) 
 and therefore probability of coincidence of BP's encounters with different  
atoms is small. Seemingly,  this allows transformation of each of the integrations \,$\int_k$\,    
in (\ref{is}) into separate collision integral:    
\begin{align} 
- \grp \! \int_k f_k  \os_{k}(\tau) \, \dots \, G_k 
\, \Rightarrow\,  \label{is0}   %  \\   \Rightarrow\,  \nonumber  
 \widehat{K} \, \os_{0}(\tau) \,\dots \, \, , 
\end{align}
so that the sum turns into solution of BLE  (\ref{ble}),  
that is   \,$\exp{[\,t\,(-V\nabla + \nu \widehat{K})]} F_0|_{t=0}$\,.  
However, such hopes are mathematically inconsistent. 
 
The matter is that characteristics of some collision in (\ref{ci}),
- namely, its relative velocity  \,$u=v-V$\,  and rigidly connected to \,$u$\,    
its impact parameter \,$b \hm = \rho -u\,(u\cdot\rho)/|u|^2$\,, -   
in reality appear to be dependent on other collisions. 
It becomes visible when one rewrites operators \,$\ol_k$\, in  (\ref{is})   
in BP's frame via  \,$u_j=v_j-V$\,: 
 \begin{align} 
 \ol_k = - u_k\nabla_{\! \rho_k}  + f_k\,(\, \frac 1\mu \nabla_{\! u_k} +   
 \frac 1M \sum_{j \neq k} \nabla_{\! u_j} -\grp )  \, ,  \label{lk}     
 \end{align} 
where \,$\mu=mM/(m+M)$\, is reduced mass of BP-atom pair. 
 From this it is clear also that the inter-dependence of collisions is as strong   
 as large is ratio  \,$\alpha =\mu/M \hm = m/(m+M)$\,,  and vanishes only 
 in the limit of infinitely massive BP. But it is obviously  insensitive to GP value 
 since GP does not enter into integrals of  (\ref{is}).   
  
 From viewpoint of mechanics, the inter-dependence  means merely that   
 chains of  \,$n-1$\, BP-atom collisions represented by \,$n$\,-th term of  (\ref{is})     
 are objects of  \,$n$\,-body problem and as such in general are wittingly irreducible   
to two-body problem. 
 A volume occupied by these chains in  \,$6(n-1)$\,-dimensional phase space  
 of  \,$n-1>1$\, atoms is not equal to product of \,$6$\,-dimensional volumes  
 of separately considered pair collisions. 
 Therefore the place of  (\ref{is0})  must be occupied by approximation   
 \begin{align} 
- \grp \! \int_k f_k  \os_{k}(\tau) \, \dots \, G_k 
\, \Rightarrow\,  \label{isa}   %  \\   \Rightarrow\,  \nonumber  
\xi_k\,  \widehat{K} \, \os_{0}(\tau) \,\dots \,  
\end{align}
with coefficients  \,$\xi_k=\xi_k(\alpha,\dots)$\,  reflecting actual        
phase volumes of collisions treated as links of coherent  many-particle events.  

{\bf 4}.  From viewpoint of randomness of the events and DFs \,$F_n$\, ($n>0$), 
 the kinematic inter-dependences between collisions look like 
 inter-particle statistical correlations. One can say that they 
 are caused by competition of atoms for their encounter with BP. 
 
Behavior of these correlations in configuration space can be qualitatively   
estimated   \cite{0710} by factor  \,$\hm\propto\,a^2/|\rho_j|^2$\,, 
 that is they decay with growing BP-atom distances like 
solid angle in which BP-atom interaction region is seen.  
Thus, under BGL the correlations do not disappear, 
but instead are self-similarly rescaled,  so that owing to \,$\nu \propto 1/a^2$\, 
a ball   \,$|\rho| < l$\,  as before receives  \,$\propto 4\pi a^2 l\nu \hm \sim l/\lambda$\, 
atoms essentially correlated with BP.  
To be more precise,  literally this concerns separate terms  
of   \,$F_n$\,'s series expansions analogous to (\ref{is}). 
But after their summation the factors  \,$\propto a^2/|\rho|^2$\,  become cut-off    
at  \,$|\rho| \gtrsim \lambda$\,. 

Hence, correlations of BP with any marked atoms are destroyed far from it    
and localized in space because of BP's interaction with the rest of gas   
\cite{tmf,pufn,lufn,0803,1203,1209,1105}. 
    On the other hand,  any correlations, once burned in a non-equilibrium process,  
  then never disappear in time, since should conserve, - 
  as required by the  ``generalized fluctuation-dissipation relations''   
 \cite{pufn}, - information necessary for time reversal of (ensemble of) 
 phase trajectories of the system.    
  
The latter statements can be formulated like a theorem \cite{0803} and imply   
 conclusion about fallacy of BLE, that is hypothesis (\ref{is0}), irrespectively to  
 GP smallness. For the proof it is  convenient to attract the coming from  \cite{0710} 
 exact non-equilibrium  ``dynamical virial relations'' (DVR)   
 \cite{tmf,0806,0803,1203,1209,1105}. 
   
 {\bf 5}. For BP in ideal gas, the DVR are easy derivable directly  
from equations  (\ref{bge}) \cite{ig}.     
In one of their equivalent forms \cite{1209}, under our initial conditions,  that are    
 \begin{align} 
 \partial_\nu  F_n = \int_{n+1} [\,F_{n+1} - G_{n+1}\,F_n\,] \, .   \label{vr0}     
 \end{align} 
We shall apply the first of them to analysis of relaxation of BP's velocity. 
With this purpose let us integrate it over BP's coordinates, thus passing to DFs 
 \,$F_0(t,V) =\int_R F_0$\,  and \,$F_1(t,V,\rho_1,v_1) \hm = \int_R F_1 \,$\,.  
Besides, on the right-hand side perform integration over \,$v_1$\,  and pass to  
\,$F_1(t,V,\rho) \hm = \int_v F_1(t,V,\rho,v)  \,$\,. Then multiply  both sides 
by \,$\nu$\, and write out the result in the form     
  \begin{align} 
 \nu \partial_\nu  F_0 = F_0 \int_{\rho} [\,\oln(t,\rho |V) -\nu\,]  =  
F_0\, \varDelta N(t|V)   \, ,   \label{vr}     
 \end{align} 
  where \,$\,\oln(t,\rho |V)  =\nu  F_1(t,V,\rho) /F_0(t,V)$\,  
  has clear meaning as conditional average value of concentration of atoms 
  at distance \,$\rho$\, from BP under condition that its velocity is known. 
 Correspondingly,   \,$\varDelta N (t | V)$\, is conditional average value 
 of fluctuations in number of atoms surrounding BP \cite{1209,1105},  
 i.e.  lack or excess of atoms around BP due to their correlations with it. 
   
 Notice, however, that the condition  ``$V$''  insufficiently characterizes  
 system's non-equilibrium as the origin of discussed correlations  
 constantly  weakening in the course of  \,$F_0(t,V)$\,'s relaxation 
 to a stationary distribution \,$F^{\infty}(V)$\, 
  (e.g. equilibrium Maxwellian one,  if  \,$G(v)$\, is such). 
  Therefore it is will be better to introduce averaging under condition 
  of realization of a given mode of relaxation,  to be represented by some 
  velocity function \,$\Psi(V)$\, orthogonal to \,$F^{\infty}$\,:  
  \,$\int_V \Psi(V)\, F^{\infty}(V) \hm = 0$\,. 
  Concretely,      
  \begin{align} 
  \varDelta N \{t | \Psi \} = \frac {\langle\,\Psi(V)\, \varDelta N (t | V) \, \rangle_t}  
  {\langle\,  \Psi(V) \, \rangle_t}   \, ,    \label{ca}     
 \end{align} 
 where \,\,$\langle \,\dots\,\rangle_t = \int_V \dots\, F_0(t,V)$\,. Then (\ref{vr})  produces   
  \begin{align} 
 \nu \partial_\nu  \,\ln\, |\,\langle\,  \Psi(V) \, \rangle_t\,|  = \varDelta N \{t | \Psi \}   \, . \label{vrm}     
 \end{align} 
 The orthogonality helps to guarantee that the denominator in (\ref{ca})   
 does not turn to zero,  although tends to it, during all time of relaxation   
(i.e. formally ever).  
As  the result, relation (\ref{vrm})   reveals transparent connection     
between a law of relaxation and integral value of accompanying  
correlations BP-gas.  
 
 {\bf 6}.  At (infinitely) small GP,  or in BGL,  stationary distribution of BP's velocity  
 undoubtedly satisfies BLE: \,$\ok F^{\infty}=0$\,, and is independent on \,$\nu$\,.  
   
  Let us suppose that BLE determines also evolution to the stationary state  
  from arbitrary initial one:  \,$\partial_t F_0 = \nu\ok F_0$\,.  
    Respectively,  choose for the role of   \,$\Psi(V)$\,  some of non-stationary   
    eigen modes of BLO, that is solutions of equation    
    \,$\ok F^{\infty} \,\Psi \hm = K F^{\infty}\,\Psi $\,  
  with non-zero  (hence, negative) eigen-values \,$K<0$\, \cite{rl}   
  (or, equivalently, \,$\ok^\dagger \Psi \hm = K\Psi$\,, where \,$\ok^\dagger$\,   
  is transposed BLO).    
  Then, according to BLE,  \,$\langle\,\Psi(V) \,\rangle_t \hm \propto E(t\nu K)$\,    
 with   exponential function  \,$E(z) \hm =\exp{(z)}$\,.    
  
Inserting this into DVR (\ref{vrm}), we get  
\,$\varDelta N \{t | \Psi \} \hm = t \nu K$\,, 
which says that number of atoms involved into correlations with BP 
unrestrictedly grows with time. 
Thus, we came  to conflict with the above underlined  finiteness 
of  spatial scales of correlations.  
Consequently, the exponential relaxation law dictated by BLE (or even its 
exponential  asymptotic) can not be realized in rigorous enough statistical mechanics.  
    
 {\bf 7}.  Just mentioned contradiction can be removed already within the approximation   (\ref{isa})  
 which suggests relaxation by law  \,$F_0(t,V) \hm = E(t\nu\ok) \,F_0(0,V)$\, and   
 \,$\langle\,\Psi(V) \,\rangle_t \hm \propto E(t\nu K)$\, with non-exponential function  
  \begin{align} 
 E(z) =  1 + \sum_{n=1}^\infty  \frac  {z^n}{n!} \, 
 \prod_{k=1}^n \xi_k \, .    \label{eo}     
 \end{align}  
  Now, relation (\ref{vrm})  implies   
  \,$\varDelta N \{t | \Psi \}  \hm = z \partial_z  \,\ln\,E(z)$\,  (\,$z=t\nu K$\,).  
  This clearly shows that the expectation of boundedness of correlation integrals,  
  i.e. the quantity \,$\varDelta N \{t | \Psi\}$\, here, justifies when \,$E(z)$\, possesses   
asymptotic of power-law type at \,$z\rightarrow -\infty$\,, for instance, simply   
\,$E(z) \propto  1/(-z)^{1+\eta}$\, with \,$\eta >0$\,. 

If it is so,   then  it looks natural to associate  limit 
\,$\varDelta N \{t\rightarrow \infty | \Psi\} \hm = -(1+\eta)$\, 
with quantity \,$-\varDelta N_-$\, from \cite{1209} 
which characterized relaxation of BP's coordinate distribution \cite{lufn},  
and its estimate there prompts us that possibly \,$\eta \hm = M/m $\,. 
Next, we want to show that the same number can be extracted in other way  
from examination of the coefficients \,$\xi_k$\,. 

{\bf 8}.  Notice that the element \,$d^3p\,d^2b\,|v-V|dt$\, of collision phase volume     
in BLO (\ref{ci}) in essence is \,$d^3p\,d^2b\,|dl|$\,, where \,$dl$\,   
means differential of atom's path relative to BP along their ``collision cylinder'' \cite{rl}. 
But in a chain of  \,$n>1$\,  collisions a ``cylinder'' of anyone of them 
occurs ``broken'' and ``spread'' because of others, thus losing its sense. 
Therefore, we are forced to define \,$dl$\,  in an inertial reference frame 
which is common for all particles taking part in the chain, 
that is in the frame pinned to their center of mass   
\,$R_{n} \hm = (MR +m\sum_{j=1}^n r_j) /M_n \hm =  
R + (m/M_n) \sum_{j=1}^n \rho_j $\,, where \,$M_n \hm =M+nm$\,. 
There, a change in position of one or another atom      
 is partly neutralized by induced displacement of the coordinate origin    
\,$R_{n}$\,,  so that in place of differential  \,$d(r_k-R) \hm = d\rho_k \hm = u_k \,dt$\,   
we find  
\,$d(r_k - R_{n}) \hm = (1-m/M_n)\, d\rho_k \hm = (M_{n-1}/M_n)\, u_k \,dt$\,. 
 
Hence, correspondingly,  in place of  \,$dl$\, in the phase volume element 
such the quantity suggests itself as      
 \,$dl_k \hm = c\,d(r_k - R_{n}) \hm = c\, (M_{n-1}/M_n)\,u_k\, dt$\,,  
where calibrating multiplier  \,$c \hm =M_1/M$\, is necessary for ensuring    
that in case of single collision (at \,$n=1$\,) one comes to the value from    
(\ref{ci}), with \,$dl/dt \hm = u$\,. 
In application to  (\ref{isa}) we must assign \,$n=k$\, (full number     
of atoms in field of vision under integral  \,$\int_k$\,).  
 In such way we obtain \,$\xi_k \hm = M_1M_{k-1}/M_kM$\,.  

Of course,  this estimate does not pretend to numeric exactness. But this 
in no way prevents its qualitative validity.  At that, it can be improved,   
taking into account that BLO (\ref{ci})  ignores  important  kinematic  
details of collision,  as if during it BP was free of a  ``kickback'' 
and changed velocity stepwise after all. Let us try to compensate 
 a waste from this distortion of mechanics, noticing that its embodiment would   
 result in true BP's velocity change only if it was supplied by 
 distorting \,$(M+m)/M$\,  times either BP's mass, up to effective value \,$M+m$\,,   
 or oppositely atom's mass, down to \,$\mu$\,. 
This correction yields  \,$\,\xi_k \hm = (1+\alpha) [1+(k-1) \alpha]/(1+k\alpha)$\,.

 {\bf 9}. Consequently, the function (\ref{eo}) appears to be   
  \begin{align} 
 E(z) =  \sum_{n=0}^\infty  \frac  {z^n}{n!} \, \frac  { (1+\alpha)^n}{1+n\alpha}   =  
 \int_0^\infty e^{\gamma z}\, w_\alpha(\gamma) \, d\gamma \, ,    \label{ef}     
 \end{align}  
where sum of the series is represented by an integral.  
Clearly,   \,$\gamma$\,  there plays role of  (dimensionless)  random rate    
of BP's  velocity relaxation,  or random probability of BP's collisions,
while  \,$w_\alpha(\gamma) $\, is probability density distribution of  
\,$\gamma$\,. It is not hard to verify that \,$w_\alpha(\gamma) $\,  is concentrated   
on interval \,$\gamma < 1+\alpha$\, where  
 \,$w_\alpha(\gamma) \hm \propto \gamma^{1/\alpha -1} \hm = \gamma^{\eta}$\, 
 with \,$\eta = M/m$\,. This agrees with conclusions obtained in  \cite{1311} 
 by methods of  ``quantum field theory in phase space'' \cite{0806}. 
 The corresponding  power-law asymptotic of \,$E(z \rightarrow -\infty)$\,, 
  i.e. that of velocity relaxation law, confirms the supposition made above. 
  
 In more detail,  if  \,$\eta \gg 1$\,, then at \,$-z =x \hm \lesssim \eta$\, 
 the relaxation is almost exponential, \,$E(-x) \approx e^{-x}$\,,  
 while at  \,$x \gtrsim \eta$\, it shows crossover to power-law ``tail'', 
 \,$E(-x) \approx \Gamma(1+\eta)/x^{1+\eta}$\,.
If  \,$\eta \lesssim 1$\,, then the relaxation law everywhere   
has more or less power-law behavior.   
 
Transforming  randomness of the velocity relaxation rate \,$\gamma$\,   
into randomness of rate (diffusivity) of BP's ``Brownian motion'',  
, \,$D \propto 1/\gamma$\,, one can find agreement with results of  
 \cite{i1,mbm} and \cite{tmf,pufn,lufn,0803,1209}. 

Naturally,  all that may be addressed to underlying qualitative difference   
of  actual statistics of BP's collisions from the Poisson type statistics  
inherent to Boltzmann's molecular chaos. In reality, in present approximation, 
according to (\ref{ef}),  probability of detection of \,$n$\, collisions  
equals to 
\,$\int_\gamma e^{-\gamma \overline{n}} (\gamma \overline{n})^n \,  
w_\alpha(\gamma) /n!$\,,  
where \,$\overline{n} \propto t$\, is average number of collisions during 
observation time.  
At  \,$\overline{n}  \hm > n$\, this probability  decreases with time by law  
 \,$\propto 1/t^{1+\eta}$\, instead of exponential one.  
 As the consequence, variance of number of collisions  grows as   
\,$ \overline{n} + \alpha^2 \overline{n} ^2/(1 + 2\alpha)$\,, 
thus violating the ``law of large numbers''. 

{\bf 10}. In conclusion, let us comment obvious defect of displayed statistical  
picture, namely,  ``quasi-static'' character of randomness of relaxation rate, 
manifesting  itself in absence of time argument  
 of \,$w_\alpha(\gamma)$\,. 
This is usual shortage of approximate approaches to solution of infinite 
BBGKY hierarchies \cite{lufn}. Its removal in mathematically more 
developed theory would lead, in particular, to appearance of effective time 
dependences of  \,$\xi_k$\, in (\ref{isa}) and transition from 
quasi-static randomness to flicker fluctuations (1/f noise). 

In principle, however,  most important thing for us here is difference   
of \,$\xi_k$\, from unit at \,$k>1$\,. May be, modern technique of computer calculations  
gives possibility to verify this at least for the third,  ``two-collision'', term  
of (\ref{is}). Then, from consideration of  problem of three bodies only one could   
obtain more strong evidence of BE's invalidity than from numeric   
``molecular dynamics''  of very many particles. 

\,\,\, 
 
  To resume, we considered velocity relaxation law for a particle interacting with   
  atoms of ideal gas, and demonstrated that actual kinematics of the interaction, 
  regardless of gas rarefaction,  forbids exponential relaxation and instead 
   establishes  one possessing power-law asymptotic, with exponent 
   determined by particle to atom mass ratio.

    \,\,\,


\begin{thebibliography}{99}

\bibitem{rl} 
P. Resibois, M. De Leener.  {\sl Classical kinetic theory of fluids}.  Wiley-Interscience, 1977. 

\bibitem{bog}
N.\,N. Bogolyubov. {\sl  Problems of dynamical theory in statistical physics}. North-Holland, 1962.  

\bibitem{i1}
Yu.\,E. Kuzovlev,  Sov. Phys. - JETP  {\bf 67} (12) 2469 (1988); 
arXiv:0907.3475 

\bibitem{ufn}
G.\,N. Bochkov, Yu.\,E.  Kuzovlev,   {\sl Sov. Phys. Uspekhi} {\bf  26} 829 (1983).  

\bibitem{tmf}
Yu.\,E. Kuzovlev,  {\sl Theor. Math. Phys.}  {\bf 160} 1301 (2009); 
arXiv:0908.0274 

\bibitem{pufn}
G.\,N. Bochkov, Yu.\,E.  Kuzovlev,   {\sl Physics - Uspekhi}  {\bf  56} (6)  590  (2013); 
arXiv:1208.1202 

  \bibitem{lufn}
Yu.\,E.  Kuzovlev,   {\sl Physics - Uspekhi}   {\bf  58} (6)  590  (2015; 
arXiv:1504.05859  
  
  \bibitem{lan}
  O.E. Lanford III, in {\sl Non-equilibrium Phenomena. 
I. The Boltzmann equation} (eds. E.W.Montroll and J.L.Lebowitz). North-Holland, 1983. 

  \bibitem{sr} 
  I. Gallagher et al.,  arXiv:1208.5753   
  
 \bibitem{kr}
N.\,S. Krylov.  {\sl Works on the foundations of statistical physics}. Princeton, 1979. 
  
  \bibitem{ct} 
  G. Calucci, D. Treleani,  arXiv:0907.4772   
    
  \bibitem{ww} 
C.-Y. Wong et al., arXiv:1505.02022    

  \bibitem{0710} Yu.E. Kuzovlev,  arXiv:0710.3831       

\bibitem{0806}   Yu.E. Kuzovlev,  arXiv:0806.4157    

\bibitem{0803}   Yu.E. Kuzovlev,   arXiv:0803.0301    

\bibitem{1203}   Yu.E. Kuzovlev,  arXiv:1203.3861       

\bibitem{1209}   Yu.E. Kuzovlev,  arXiv:1209.5425   

\bibitem{1105}   Yu.E. Kuzovlev,   arXiv:1105.0025     

\bibitem{ig}   Yu.E. Kuzovlev,  arXiv:0902.2855\,,   arXiv:0911.0651 

\bibitem{mbm}   Yu.E. Kuzovlev,  cond-mat/0609515        
\,,\, arXiv:1007.1992 

\bibitem{1311}   Yu.E. Kuzovlev,  arXiv:1311.3152        

\end{thebibliography}
\end{document}